\documentclass[twocolumn,conference]{IEEEtran}
\IEEEoverridecommandlockouts
\normalsize
\usepackage{fixltx2e}
\usepackage{mathrsfs}
\usepackage{amsmath}
\usepackage{amssymb}
\usepackage{mathtools}
\usepackage{graphics}
\usepackage{float}
\usepackage{epstopdf}
\usepackage{graphicx}
\usepackage{bm}
\usepackage{tikz}
\usepackage[lined,boxed,commentsnumbered,linesnumbered,ruled]{algorithm2e}
\usepackage{cite}
\usepackage{comment}
\usepackage{balance}
\usepackage{booktabs} % required for showing good \hline in matrix form --> \midrule
\usepackage{enumitem}
\usepackage{mathtools}
\usetikzlibrary{shapes, patterns,decorations.text, decorations.pathreplacing}
\usepackage{ifthen}
\usepackage{multirow}
\usepackage[capitalize]{cleveref}
\crefname{equation}{\unskip}{\unskip}
\newcommand*{\Scale}[2][4]{\scalebox{#1}{\ensuremath{#2}}}%

\newtheorem{example}{Example}
\newtheorem{theorem}{Theorem}
\newtheorem{definition}{Definition}
\newtheorem{lemma}{Lemma}

\allowdisplaybreaks
\newcommand{\X}{\bm X}

\newcommand{\eqdef}{\triangleq}              % definition
\newcommand{\set}[1]{\mathcal{#1}}           % set
\newcommand{\mat}[1]{\bm{#1}}                % matrices
\newcommand{\GF}{\mathrm{GF}}                % Galois-field
\newcommand{\Nat}[1]{\mathbb{N}_{#1}}        % natural numbers until N
\newcommand{\code}[1]{\mathcal{#1}}          % the code notation
\newcommand{\const}[1]{\textnormal{\usefont{U}{eur}{m}{n}\selectfont #1}} % Euler
\newcommand{\rank}[1]{\mathrm{rank}\left(#1\right)} % rank operator

\newcommand{\bigrank}[1]{\operatorname{rank}\bigl(#1\bigr)}

\newcommand{\modify}[1]{\textcolor{black}{#1}}
  
\renewcommand{\r}{\color{red}}

\makeatletter
\newcommand*\rel@kern[1]{\kern#1\dimexpr\macc@kerna}
\newcommand*\widebar[1]{%
  \begingroup
  \def\mathaccent##1##2{%
    \rel@kern{0.8}%
    \overline{\rel@kern{-0.8}\macc@nucleus\rel@kern{0.2}}%
    \rel@kern{-0.2}%
  }%
  \macc@depth\@ne
  \let\math@bgroup\@empty \let\math@egroup\macc@set@skewchar
  \mathsurround\z@ \frozen@everymath{\mathgroup\macc@group\relax}%
  \macc@set@skewchar\relax
  \let\mathaccentV\macc@nested@a
  \macc@nested@a\relax111{#1}%
  \endgroup
}
\makeatother

\begin{document}

\title{Local Reconstruction Codes: A Class of MDS-PIR Capacity-Achieving Codes} %
\author{\IEEEauthorblockN{Siddhartha Kumar\IEEEauthorrefmark{2}, Hsuan-Yin Lin\IEEEauthorrefmark{2},  Eirik Rosnes\IEEEauthorrefmark{2}, and Alexandre Graell i Amat\IEEEauthorrefmark{3}}\\%[1ex]
\vspace{-4mm}\IEEEauthorblockA{\IEEEauthorrefmark{2}Simula UiB, N--5020 Bergen, Norway}
\IEEEauthorblockA{\IEEEauthorrefmark{3}Department of Electrical Engineering, Chalmers University of Technology, SE--41296 Gothenburg, Sweden%\\email: \{kumarsi, eirikrosnes\}@simula.no, alexandre.graell@chalmers.se
}
\thanks{The work of S. Kumar, H.-Y. Lin, and E.\ Rosnes was partially funded by the Research Council of Norway (grant 240985/F20). A.\ Graell i Amat was supported by the Swedish Research Council under grant \#2016-04253.}}%
%\thanks{This work was partially funded by the Research Council of Norway (grant 240985/F20) and the Swedish Research Council (grant
   % \#2016-04253).}}%

\maketitle

\begin{abstract}

%The efficiency of a private information retrieval (PIR) protocol for a distributed storage system (DSS) is characterized by its PIR rate, which also depends on the underlying storage code. In the classical case when no nodes in the DSS are allowed to collude, it was shown recently that the maximum distance separable (MDS)-PIR capacity, i.e., the maximum achievable PIR rate by any protocol where the  underlying code is MDS, can be achieved even by non-MDS codes. We refer to codes that achieve this capacity as MDS-PIR capacity-achieving codes.  

We prove that a class of distance-optimal local reconstruction codes (LRCs), an important family of repair-efficient codes for distributed storage systems, achieve the maximum distance separable private information retrieval capacity for the case of noncolluding nodes. This particular class of codes includes Pyramid codes and other LRCs proposed in the literature.
\end{abstract}

\section{Introduction}

Private information retrieval (PIR) deals with the scenario where a user wants to retrieve a data item from a database without letting the database know the identity of the requested item. PIR was first introduced in the computer science literature by Chor \emph{et al.} in \cite{cho95}, where the authors considered that the database is replicated across $n$ servers (nodes) and presented a PIR protocol 
that efficiently achieves privacy in the presence of a single spy node. In \cite{cho95}, the efficiency of the PIR protocol was measured in terms of upload and download cost. 

With the advent of distributed storage systems (DSSs), where data is stored in a distributed fashion over a number of nodes using a storage code rather than simply replicated, the concept of PIR has gained traction in the information theory community. As typically the size of the data items stored is much larger compared to the size of the queries sent to the nodes, the upload cost is negligible compared to the download cost \cite{cha15}. Thus, under the information-theoretic formulation, the efficiency of a PIR protocol, referred to as the PIR rate, is measured in terms of download cost. More precisely, the PIR rate is defined as the ratio between the requested file size and the total amount of downloaded data. The maximum PIR rate over all PIR protocols is the PIR capacity.

The authors in \cite{sha14} were the first to introduce PIR protocols for DSSs in the information theory community, assuming that data is stored using two explicit linear codes. In \cite{cha15}, an upper bound on the PIR rate for a certain class of linear PIR protocols was given. 
%\textcolor{blue}{In \cite{Faz15}, the authors showed that any $n$-server PIR protocol can be efficiently emulated while preserving privacy and the overall communication cost when the DSS uses a certain coding scheme (referred to as PIR codes).} \textcolor{red}{Comment: I assume that you realize that the sentence in blue makes little sense and no one will understand. What do you mean by ``a protocol can be emulated''?? Honestly, let's be rigorous! We can't simply write an extremely vague sentence and be happy with it. I suggest to simply remove it. I understand that you want to please Eitan, but we are considering another kind of PIR here.} 
For the case of replicated data and a single spy node, commonly known as the noncolluding case, Sun and Jafar \cite{SunJafar17_1} derived the PIR capacity and  presented a PIR capacity-achieving scheme. Also, for the noncolluding case, Banawan and Ulukus \cite{BanawanUlukus18_1} derived the maximum achievable PIR rate for the more general scenario where data is stored in the DSS using a maximum distance separable (MDS) code and presented a scheme that achieves it. As the underlying storage code is an MDS code, such a maximum achievable PIR rate is usually referred to as the MDS-PIR capacity.

The MDS-PIR capacity depends on the code rate of the underlying MDS storage code and the number of files stored in the DSS. In \cite{taj16}, a PIR protocol for MDS-coded data that achieves the \emph{asymptotic} MDS-PIR capacity when the number of files tends to infinity was presented. In \cite{Kum17b}, the authors presented a PIR  protocol for the case where the underlying storage code can be an arbitrary linear code and numerically showed that the proposed protocol can achieve the asymptotic MDS-PIR capacity even if the underlying storage code is non-MDS. With some abuse of language,  we refer to such codes as MDS-PIR capacity-achieving codes. While the aformentioned protocols assume that nodes in the DSS do not collude,  \cite{FreijHollantiGnilkeHollantiKarpuk17_1,FreijHollantiGnilkeHollanti17_1sub, KumarLinRosnesGraell17_1sub,taj18} proposed PIR schemes for the case of colluding nodes. %However, the PIR capacity for the colluding case is not known.

In a DSS, the storage code is used not just to achieve reliability against node failures, but also to repair failed nodes. 
%It is quite well known that MDS codes are not efficient in repairing node failures.
Although MDS codes are optimal in terms of storage overhead (for a given rate), they are characterized by a large repair locality, i.e., the repair of a failed node requires contacting a large number of nodes.
%a high node repair cost. 
Thus, with  focus on repair locality, several code constructions such as Pyramid codes \cite{Hua07}, locally repairable codes \cite{Sat13}, and local reconstruction codes (LRCs) \cite{Hua12} have been proposed. Such codes follow a similar design philosophy, and we refer to them globally as LRCs. In \cite{Kum17b}, it was shown numerically that, interestingly, the asymptotic MDS-PIR capacity for the case of noncolluding nodes can be achieved for some Pyramid codes. 

In this paper, we go a step further and formally prove that an important class of repair-efficient storage codes, namely a class of distance-optimal LRCs, are MDS-PIR capacity-achieving codes in the noncolluding case. This implies that one does not need to sacrifice on  the repair locality to achieve the MDS-PIR capacity. 

\section{Definitions and Preliminaries}

Throughout the paper we use the following notation. We represent the set of $a$ consecutive integers as $\Nat{a}\triangleq\{1,\ldots,a\}$, while $\Nat{a:b}\triangleq\{a,\ldots,b\}$ represents the set of integers from $a$ to $b$. We use calligraphic upper case, bold upper case, and bold lower case letters to denote sets, matrices, and vectors, respectively.  As an example, $\mathcal X$, $\bm X$, and $\bm x$ represent a set, matrix, and a vector, respectively. The identity matrix of order $a$ is denoted by $\bm I_a$, and  %, and the all-zero and all-one matrices of dimensions $a\times b$ are denoted by $\bm 0_{a\times b}$ and $\bm 1_{a\times b}$, respectively. 
$(\bm X_1|\ldots|\bm X_a)$ denotes the horizontal concatenation of matrices $\bm X_1,\ldots,\bm X_a$.
A submatrix of $\bm X$ that is restricted in columns by the set $\mathcal J$ is denoted by $\bm X|_{\mathcal J}$, and the rank of $\bm X$ is denoted by $\rank{\bm X}$. $\mathcal C$ denotes an $[n,k]$ linear code of block length $n$, dimension $k$, and minimum Hamming distance $d_{\mathsf{min}}^{\mathcal C}$  over the Galois field GF$(q)$. A generator matrix of $\mathcal C$ is denoted by $\bm G^{\mathcal{C}}$, while $\bm H^{\mathcal{C}}$ denotes a parity-check matrix. $\mathcal C|_{\mathcal J}$ is the punctured code obtained from $\mathcal C$ by restricting the code coordinates to the indices in $\mathcal J$.  %The support of a vector $\vect{x}$, i.e., the coordinate set of nonzero entires of $\vect{x}$, is denoted by $\chi(\vect{x})$. 
A set of coordinates of $\mathcal{C}$, $\set{J}\subseteq\Nat{n}$, of size $k$ is said to be an \emph{information set} if and
only if $\mat{G}^\code{C}|_\set{J}$ is invertible. With some abuse of language, we sometimes interchangeably refer to binary vectors as erasure patterns under the implicit assumption that the ones represent erasures.
%
%An erasure pattern (or binary vector) $\bm{x}$ is said to be correctable by a code $\code{C}$ if
  %matrix $\bm H^{\code{C}}|_{\chi(\bm{x})}$ has rank $|\chi(\bm{x})|$. 

%\subsection{System Model}

We consider a DSS that stores $f$ files $\bm X^{(1)},\ldots, \bm X^{(f)}$, where 
$\bm{X}^{(m)}=(x_{i,j}^{(m)})$, $m\in\Nat{f}$, can be seen as a $\beta \times k$ matrix over $\GF(q^\ell)$,
with $\beta, k, \ell\in\Nat{}$. Let $\bm x^{(m)}_i$ denote the $i$-th row of $\X^{(m)}$. Each $\bm x_i^{(m)}$ is encoded by an $[n,k]$ code $\mathcal{C}$ over
$\GF(q)$ into a length-$n$ codeword
$\bm c^{(m)}_i=\bigl(c^{(m)}_{i,1},\ldots,c^{(m)}_{i,n}\bigr)$, where $c_{i,j}^{(m)}\in\GF(q^\ell)$, $j\in\Nat{n}$, is stored on the $j$-th node. The symbols are stored in the order of increasing $m$ and secondly in the order of increasing $i$ (see \cite[Sec.~III]{KumarLinRosnesGraell17_1sub}).

\subsection{MDS-PIR Capacity-Achieving Codes}
\label{sec:CapAch}

%The MDS-PIR capacity was derived in \cite{BanawanUlukus18_1} and is given in the following theorem.
%\begin{theorem}
For a given number of files $f$ stored using an $[n,k]$ MDS code, the MDS-PIR capacity \cite[Thm.~1]{BanawanUlukus18_1}  is 
$\const{C}_f = \frac{1-k/n}{1-(k/n)^f}$. We refer to $\const{C}_f$ as the \emph{finite} MDS-PIR capacity, as it depends on the number of files. 
When the number of files grows very large, i.e., $f\rightarrow\infty$, the MDS-PIR capacity reduces to
%\begin{align*}
$\const{C}_\infty = 1-\frac{k}{n}$,
%\end{align*}		
which we refer to as the \emph{asymptotic} MDS-PIR capacity.

We denote by $\const{R}_f(\code{C})$ the PIR rate of a PIR scheme that uses code $\mathcal C$ as the underlying storage code to store $f$ files.
The following theorem gives a condition for the existence of MDS-PIR capacity-achieving codes (under Protocols 1 and 2 presented by the authors in \cite{KumarLinRosnesGraell17_1sub}).\footnote{Protocol~2 in \cite{KumarLinRosnesGraell17_1sub} was originally introduced in \cite{Kum17b}.} 
\begin{theorem}\label{Th:CapAch}
 Consider a DSS that uses an $[n,k]$ code $\mathcal C$ to store $f$ files. If there exists a binary $n\times n$ matrix $\bm E$ of  row and column weight $n-k$ such that each row is an erasure pattern that is correctable  by $\mathcal C$, then $\mathcal C$ achieves the finite MDS-PIR capacity $\const{C}_f$ (under Protocol~1 in \cite{KumarLinRosnesGraell17_1sub}), i.e.,
 %\begin{align*}  %\label{eq:PIRachievable-rate_code}
    $\const{R}_f(\code{C}) = \const{C}_f$,
%\end{align*}
and the asymptotic MDS-PIR capacity $\const{C}_\infty$ (under Protocol~2 in \cite{KumarLinRosnesGraell17_1sub}), i.e.,
 %\begin{align*}
    $\const{R}_\infty(\code{C}) = \const{C}_\infty$.
%\end{align*}
\end{theorem}

In Sections~\ref{Sec: Ematrix_LRC} and \ref{sec:Sketch}, we prove that for a class of 
distance-optimal $(r,\delta)$ information locality codes \cite{Kam14}, an important class of LRCs, such an $\bm E$ exists, and hence this class of codes is MDS-PIR capacity-achieving.

\subsection{Local Reconstruction Codes}
\label{sec:local-reconstr-codes}

LRCs are a family of codes characterized by their low repair locality, i.e., in order to repair a failed node, only a relatively low number of nodes need to be contacted. 
In particular, we consider \emph{information locality} codes, which are
systematic codes whose focus is to reduce the repair locality of systematic nodes (i.e., nodes that store systematic code symbols) \cite{Hua07,Hua12,Sat13,Kam14}. 
%On the contrary, LRCs that achieve low locality for all nodes are referred to as \emph{all-symbol locality} codes. The codes presented in \cite{tam14} are examples of all-symbol locality codes. 
Formally, they are defined as follows.
\begin{definition}[{$(r,\delta)$ information locality code\cite[Def.~2]{Kam14}}]
\label{Def: LRC}
An $[n,k]$ code $\mathcal{C}$  is said to be an $(r,\delta)$ information locality code if there exist $L_{\mathsf c}$ punctured codes $\code{C}_j\eqdef\set{C}|_{\set{S}_j}$ of $\mathcal C$ with column coordinate set $\mathcal S_j\subset\Nat{n}$ for $j\in\Nat{L_{\mathsf{c}}}$. Furthermore, $\{\code{C}|_{\set{S}_j}\}_{j\in\Nat{L_{\mathsf{c}}}}$ must satisfy the following conditions:
\begin{enumerate}
\item $|\mathcal S_j|\leq r+\delta-1$, $\forall\,j\in\Nat{L_{\mathsf{c}}}$,
\item $d_{\mathsf{min}}^{\mathcal C_j}\geq\delta$, $\forall\,j\in\Nat{L_{\mathsf{c}}}$, and
\item $\bigrank{\bm G^{\mathcal C}|_{\bigcup_j\set{S}_j}}=k$.
\end{enumerate}
\end{definition}

\modify{In other words, \cref{Def: LRC} says that there are $L_\mathsf{c}$ local codes in $\mathcal C$ each having a block length of at most $r+\delta-1$, minimum Hamming distance at least $\delta$, and the \modify{union of all coordinate sets of the local codes contains} an information set. The overall code $\mathcal C$ has $d_{\mathsf{min}}^{\mathcal C}\leq n-k+1-(\lceil k/r\rceil-1)(\delta-1)$ and can repair up to $\delta-1$ systematic nodes by contacting $r$ storage nodes. Codes that achieve the upper bound on the $d_{\mathsf{min}}$ are known as distance-optimal $(r,\delta)$ information locality codes and  have the following structure.}

\begin{definition}[Distance-optimal $(r,\delta)$ information locality code {\cite[Thm.~2.2]{Kam14}}]
\label{Def: OptLRCs}
Let $r\mid k$ such that $L_\mathsf{c}=k/r$. An $(r,\delta)$ information locality code $\mathcal C$ as defined in \cref{Def: LRC} is distance-optimal if:
\begin{enumerate}
\item Each local code $\mathcal C|_{\set{S}_j}$, $j\in\Nat{L_\mathsf{c}}$, is an $[r+\delta-1,r]$ MDS code defined by a parity-check matrix   $\mat{H}^{\mathcal C|_{\set{S}_j}}=(\bm P_j|\bm I_{\delta-1})$ of dimensions $(\delta-1)\times(r+\delta-1)$ and minimum Hamming distance $d_{\mathsf{min}}^{\mathcal C|_{\set{S}_j}}=\delta$. %The matrix   $\bm P_j$, of dimensions $(\delta-1)\times r$, is the parity-check matrix of the $i$-th local code.
\item The sets $\{\mathcal S_j\}_{j\in\Nat{L_\mathsf{c}}}$ are disjoint, i.e., $\set{S}_j\cap\set{S}_{j'}=\emptyset$ for   all $j,j'\in\Nat{L_\mathsf{c}}$, $j\not=j'$.
\item The code $\mathcal C$ has a parity-check matrix of the form
\end{enumerate}
\begin{align}
 \Scale[0.95]{\label{Eq: H_opt_LRC}
  \mat{H}=\left(\begin{array}{ccccccc|c}
    \bm P_1 & \bm I_{\delta-1} &           &                  &       &             &                  & \\
            &                  & \bm P_2   & \bm I_{\delta-1} &       &             &                  & \\
            &                  &           &                  &\ddots &             &                  & \\
            &                  &           &                  &       & \bm P_{L_\mathsf{c}} & \bm I_{\delta-1} & \\
  \hline
  \mat{M}_1 & \bm 0            & \mat{M}_2 & \bm 0            &\cdots & \mat{M}_{L_\mathsf{c}} & \bm 0 & \bm I_a\\ 
  \end{array}\right)}
\end{align}
where the matrices $\bm M_1,\ldots, \bm M_{L_\mathsf{c}}$ are arbitrary matrices in $\GF(q)$ of dimensions $(n-L_\mathsf{c}(r+\delta-1))\times r$,  and $a\eqdef n-L_\mathsf{c}(r+\delta-1)$.
\end{definition}

For ease of exposition, we refer to the local parities as the parity symbols that take part in the local codes, while the parity symbols that are not part of the $L_{\mathsf c}$ local codes are referred to as global parity symbols. According to \cref{Def: OptLRCs}, there exist $n-L_{\mathsf{c}}(r+\delta-1)$ global parities and $L_{\mathsf{c}}(\delta-1)$ local parities. We partition the coordinates of these parities into $L+1$ sets, where $L\eqdef\bigl\lfloor \frac{n}{r+\delta-1}\bigr\rfloor$. For $j\in\Nat{L+1}$, we have
\begin{align}
  \mathcal P_j
  =\begin{cases}
    \{(j-1)n_\mathsf{c}+r+1,\ldots,j n_\mathsf{c}\} & \text{if }j\in\Nat{L_{\mathsf c}},\\
    \{(j-1)n_\mathsf{c}+1,\ldots,j n_\mathsf{c}\} & \text{if }j\in\Nat{L_{\mathsf c}+1: L},\\
    \{L n_\mathsf{c}+1,\ldots,n\} & \text{if } j=L+1,
  \end{cases}\label{eq:parities_LRC}
\end{align} 
where $n_\mathsf{c}\eqdef r+\delta-1$ is the block length of each local code. The set $\mathcal P_{j}$, $j\in\Nat{L_{\mathsf c}}$, represents the coordinates of the local parities of the $j$-th local code $\mathcal C_{j}$. The remaining sets $\mathcal P_{j}$, $j\in\Nat{L_{\mathsf c}+1:L+1}$, represent the coordinates of the global parities of $\mathcal C$. As such, the set $\set{P}=\bigcup_{j=1}^{L+1} \mathcal P_j$ represents the parity coordinates of $\mathcal C$.

\section{Distance-Optimal Local Reconstruction Codes are MDS-PIR Capacity-Achieving}
\label{Sec: Ematrix_LRC}
%As shown in Section~\ref{sec:CapAch}, the only requirement for a code to be MDS-PIR capacity-achieving is that $\mat{\Lambda}_{\kappa,\nu}(\code{C})$ with $\frac{\kappa}{\nu}=\frac{k}{n}$ exists. For simplicity, we take $\kappa=k$ and $\nu=n$. Thus, for an  LRC $\mathcal C$, we need to show that $\mat{\Lambda}_{k,n}(\code{C})$ exists. Let $\bm E$ be an $n\times n$ binary matrix where each row represents an erasure pattern that is correctable by $\mathcal C$, in which the symbol $1$ represents an erasure. By the relation between correctable erasure patterns and information sets we have
%\begin{align*}
%	\bm E=\bm 1_{n\times n}+\mat{\Lambda}_{k,n}(\code{C}).
%\end{align*}
%Furthermore, by this relation, $\bm E$ is an $(n-k)$-regular matrix. 
%In the following, we show the existence of $\bm E$ for any distance-optimal LRC. 

Consider an $[n,k]$ distance-optimal $(r,\delta)$ information locality code (see \cref{Def: OptLRCs}) for which
the $(n'-k) \times n'$ matrix
  \begin{align}
  \label{Eq: H_MDS}
  \left(\begin{array}{cccc|c}
          \bm P_1 & \bm P_2 & \cdots & \bm P_{L_{\mathsf c}} & \multirow{2}*{$\bm I_{n'-k}$} \\
          \bm M_1 & \bm M_2 & \cdots & \bm M_{L_{\mathsf c}} & \\
        \end{array}\right) \triangleq \bm H^{\mathsf{MDS}}
\end{align}
is the parity-check matrix of an $[n',k]$ MDS code over $\GF(q)$, where
$n'=n-(L_\mathsf{c}-1)(\delta-1)$.\footnote{Examples of codes that satisfy \eqref{Eq: H_MDS} are Pyramid codes, the LRCs in \cite{Hua12}, and codes from the parity-splitting construction of \cite{Kam14}.} For such a class of codes, we give an explicit construction of the matrix $\bm E$ in order to design the PIR
protocol.

Recall that $L=\bigl\lfloor\frac{n}{n_\mathsf{c}}\bigr\rfloor$, $n_\mathsf{c}=r+\delta-1$, and let
$\bar r\eqdef n\bmod n_\mathsf{c}$. We consider
\begin{IEEEeqnarray*}{rCl}
  \mat{E}=
  \begin{pmatrix}
    \mat{E}_{1,1}& \mat{E}_{1,2}& \ldots& \mat{E}_{1,L+1}
    \\
    \vdots& \vdots& \vdots& \vdots
    \\
    \mat{E}_{L+1,1}& \mat{E}_{L+1,2}& \ldots& \mat{E}_{L+1,L+1}
  \end{pmatrix}
\end{IEEEeqnarray*}
having $(L+1)^2$ submatrices $\mat{E}_{l,h}$, $l,h\in\Nat{L+1}$. For any $l,h\in\Nat{L}$,
the submatrices $\mat{E}_{l,h}$ have dimensions $n_\mathsf{c}\times n_\mathsf{c}$, $\mat{E}_{l,L+1}$ has dimensions 
  $n_\mathsf{c}\times\bar{r}$, $\mat{E}_{L+1,h}$ has dimensions $\bar{r}\times n_\mathsf{c}$, and $\mat{E}_{L+1,L+1}$ has dimensions 
  $\bar{r}\times \bar{r}$. We denote by $\bm e_i^{(l)}$, $l\in\Nat{L+1}$, the $i$-th row of $\bigl(\mat{E}_{l,1}|\ldots|\mat{E}_{l,L+1}\bigr)$. The coordinates of $\bm e_i^{(l)}$ represent the coordinates of the code
$\mathcal C$ defined by its parity-check matrix in \cref{Eq: H_opt_LRC}. Furthermore, each row vector is subdivided into
$L+1$ subvectors $\bm e^{(l)}_{i,j}$, $j\in\Nat{L+1}$, as
\begin{align*}
  \bm e_i^{(l)}=(e_{i,1}^{(l)},\ldots,e_{i,n}^{(l)})
  =(\bm e_{i,1}^{(l)},\ldots, \bm e_{i,L}^{(l)},\bm e_{i,L+1}^{(l)}).
\end{align*}
The subvectors $\bm e^{(l)}_{i,1},\ldots, \bm e^{(l)}_{i,L}$ are of length $n_\mathsf{c}$, while $\bm e^{(l)}_{i,L+1}$ is of
length $\bar r$. Correspondingly, we can think about $\mat{E}$ as partitioned into $L+1$ column partitions, where the first $L_\mathsf{c}$ partitions correspond to the $L_\mathsf{c}$ local codes and the remaining $L+1-L_\mathsf{c}$ partitions correspond to global parities (see also \eqref{eq:parities_LRC}). We can write $\mat{E}$ as
\begin{align*}
\Scale[0.95]{
  \bm E\eqdef\left(\begin{matrix}
      \bm e_1^{(1)}\\
      \vdots\\
      \bm e_{n_\mathsf{c}}^{(1)}\\
      \vdots\\
      \bm e_{n_\mathsf{c}}^{(L)}\\[1mm]
      \bm e_{1}^{(L+1)}\\
      \vdots\\
      \bm e_{\bar r}^{(L+1)}
    \end{matrix}\right)=\left(\begin{matrix}
      \bm e^{(1)}_{1,1} & \bm e^{(1)}_{1,2} & \cdots & \bm e^{(1)}_{1,L} & \bm e^{(1)}_{1,L+1}\\
      \vdots  & \vdots  & \cdots & \vdots & \vdots\\
      \bm e^{(1)}_{n_\mathsf{c},1} & \bm e^{(1)}_{n_\mathsf{c},2} & \cdots & \bm e^{(1)}_{n_\mathsf{c},L} & \bm e^{(1)}_{n_\mathsf{c},L+1}\\
      \vdots  & \vdots  & \cdots & \vdots & \vdots\\
      \bm e^{(L)}_{n_\mathsf{c},1} & \bm e^{(L)}_{n_\mathsf{c},2} & \cdots & \bm e^{(L)}_{n_\mathsf{c},L} & \bm e^{(L)}_{n_\mathsf{c},L+1}\\[1mm]
      \bm e^{(L+1)}_{1,1} & \bm e^{(L+1)}_{1,2} & \cdots & \bm e^{(L+1)}_{1,L} & \bm e^{(L+1)}_{1,L+1}\\
      \vdots  & \vdots  & \cdots & \vdots & \vdots\\
      \bm e^{(L+1)}_{\bar r,1} & \bm e^{(L+1)}_{\bar r,2} & \cdots & \bm e^{(L+1)}_{\bar r,L} & \bm e^{(L+1)}_{\bar r,L+1}
    \end{matrix}\right)}.
\end{align*}
We refer to the set of rows $\bm e_1^{(l)},\ldots, \bm e_{n_\mathsf{c}}^{(l)}$ as the $l$-th row partition of $\bm E$.

For convenience, we divide $\bm E$ into four submatrices $\tilde{\bm E}$, $\bm W$, $\bm Z$, and $\bm O$ defined as
\begin{IEEEeqnarray*}{rCl}
\Scale[0.95]{
  \tilde{\mat{E}}}& \eqdef &
  \Scale[0.95]{\begin{pmatrix}
    \bm e^{(1)}_{1,1} & \bm e^{(1)}_{1,2} & \cdots & \bm e^{(1)}_{1,L}\\
    \bm e^{(1)}_{2,1} & \bm e^{(1)}_{2,2} & \cdots & \bm e^{(1)}_{2,L}\\
    \vdots  & \vdots  & \cdots & \vdots\\
    \bm e^{(L)}_{n_\mathsf{c},1} & \bm e^{(L)}_{n_\mathsf{c},2} & \cdots & \bm e^{(L)}_{n_\mathsf{c},L}
  \end{pmatrix}},
  \Scale[0.95]{\bm Z\eqdef
  \begin{pmatrix}
    \bm e^{(1)}_{1,L+1}\\
    \bm e^{(1)}_{2,L+1}\\
    \vdots\\
    \bm e^{(L)}_{n_\mathsf{c},L+1}
  \end{pmatrix}},
  \nonumber\\[1mm]
  \Scale[0.95]{\bm W}& \Scale[0.95]{\eqdef} &
  \Scale[0.95]{\begin{pmatrix}
    \bm e^{(L+1)}_{1,1} & \bm e^{(L+1)}_{1,2} & \cdots & \bm e^{(L+1)}_{1,L}\\
    \vdots  & \vdots  & \cdots & \vdots\\
    \bm e^{(L+1)}_{\bar r,1} & \bm e^{(L+1)}_{\bar r,2} & \cdots & \bm e^{(L+1)}_{\bar r,L}
  \end{pmatrix},
  \bm O\eqdef
  \begin{pmatrix}
    \bm e^{(L+1)}_{1,L+1}\\
    \vdots\\
    \bm e^{(L+1)}_{\bar r,L+1}
  \end{pmatrix}},%\IEEEeqnarraynumspace %\label{Eq: SubMats_E}
\end{IEEEeqnarray*}
where $\tilde{\bm E}$ is an $n_\mathsf{c}L\times n_\mathsf{c}L$ matrix, having $L^2$ submatrices $\mat{E}_{l,h}$,
$l,h\in\Nat{L}$.

%In the following, we give a systematic procedure for permuting columns of certain rows in $\bm E$ in order to obtain an
%$(n-k)$-regular matrix.
In the following, we give a systematic construction of $\bm E$ such that it is $(n-k)$-regular.\footnote{For ease of notation, we will refer to a matrix with constant row weight, constant column weight, and constant row and column weight equal to $a$ as an $a$-row regular, $a$-column regular, and $a$-regular matrix, respectively.}
 The construction involves two steps.
\begin{enumerate}
\item[a)] \textbf{Initialize matrices $\tilde{\bm E}$, $\bm W$, $\bm Z$, and $\bm O$.} Matrix $\bm Z$ is
  initialized to the all-zero matrix of dimensions $n_\mathsf{c}L\times \bar r$.  Matrices $\mat{W}$ and $\mat{O}$ are
    initialized by setting $e^{(L+1)}_{i,j}=1$, $i\in\Nat{\bar r}$, $j\in\set{P}=\bigcup_{j'=1}^{L+1}\set{P}_{j'}$, where $\set{P}$
    corresponds to the parity coordinates of $\code{C}$ and the sets $\set{P}_{j'}$ are defined in
    \cref{sec:local-reconstr-codes} (see \eqref{eq:parities_LRC}). Let 
    $m=\bigl\lfloor\frac{n-k}{L}\bigr\rfloor$, $m_1= m+1$, $\rho_1=\cdots=\rho_t=m_1$, and
    $\rho_{t+1}=\cdots=\rho_{L}=m$, where $t=(n-k)\bmod L$. Matrix $\tilde{\bm E}$ is initialized with the
  structure
\begin{align}
  \label{Eq: Ehat_structure}
  \tilde{\bm E}=
\left(\begin{matrix}
      \bm \pi_1  & \bm \pi_2 & \cdots & \bm \pi_L\\
      \bm \pi_L  & \bm \pi_1 & \cdots & \bm \pi_{L-1}\\
      \vdots     & \vdots    & \cdots & \vdots\\
      \bm \pi_2  & \bm \pi_3 & \cdots & \bm \pi_1
    \end{matrix}\right),
\end{align}
where each matrix entry $\bm\pi_{l}$, $l\in\Nat{L}$, is a $\rho_l$-regular square matrix of order $n_\mathsf{c}$. Notice
that due to the structure in \cref{Eq: Ehat_structure}, $\tilde{\bm E}$  has row and column weight equal to $n-k$,
and subsequently each row of $\bm E$ has weight $n-k$. Note also that the columns of $\bm E$ with coordinates in $\mathcal P_j$, $j\in\Nat{L}$, have column weight $n-k+\bar{r}$, while the columns with coordinates in $\mathcal P_{L+1}$ have weight $\bar r$.

\item[b)] \textbf{Swapping elements between $\tilde{\bm E}$ and $\bm Z$.}
\modify{The swapping of elements is performed iteratively with $\bar r$ iterations. For each iteration, in the $i$-th row partition and $j$-th column partition, we consider a set of row coordinates $\mathcal R^{(i)}_j$ of size $|\mathcal{P}_j|$ from which $s_{j}^{(i)}\in\{0,1\}$ ones from columns with coordinates in $\mathcal P_j$, $j \in \Nat{L}$,  are swapped with zeroes in the corresponding rows of $\bm Z$. For convenience, we define $\bm s^{(i)}=(s_{1}^{(i)},\ldots,s_{L}^{(i)})$ and require that  $\sum_{j=1}^{L}s^{(i)}_j = 1$.  Note that  $\mathcal R^{(i)}_j$ and $\bm s^{(i)}$ depend on the iteration number. We describe the procedure for iteration $j' \in \Nat{\bar{r}}$. For the first row partition, select $\bm s^{(1)}$ with $s^{(1)}_j=1$ and $s^{(1)}_z=0$, $\forall\, z\in\Nat{L}\backslash\{j\}$, for some $j\in\Nat{L}$, such that if $j \in \Nat{L_{\mathsf c}}$ there exist $\delta-1$ rows in the first row partition and $j$-th column partition such that their individual weight is strictly larger than $\delta-1$, and otherwise if $j \in \Nat{L_{\mathsf c}+1:L}$, all rows in the first row partition and $j$-th column partition must have weight  larger than or equal to $\max(1,m-(\delta-1))$. This will ensure that the resulting erasure patterns after the swap (as described next) are correctable by $\mathcal{C}$ (see Section~\ref{sec:Sketch}). Such an $\bm s^{(1)}$ will also always exist   for all $\bar{r}$ iterations as shown in Section~\ref{sec:Sketch} below. Next, for all $i'\in\mathcal R^{(1)}_j$ and $p\in\mathcal P_j$ (where different $p$'s are chosen for different $i'$'s, and index $j$ is such that $s^{(1)}_j=1$)} the one at coordinate  $(i',p)$ of $\tilde{\bm E}$ is swapped with a zero at coordinate $(i',j')$ of $\bm Z$  (this corresponds to coordinate $(i',n_{\mathsf c}L+j')$ of $\bm E$). Then, for the remaining row partitions $i=2,\ldots,L$, consider $\bm s^{(i)}$ to be  the $(i-1)$-th right cyclic shift of $\bm s^{(1)}$ and repeat the swapping procedure for the first row partition. %, i.e., swap the one at  coordinate $(i',p)$ with the zero at coordinate $(i',j')$ for all $i'\in\mathcal R^{(i)}$ and $p\in\mathcal P_j$. 
\modify{Due to the specific selection of $\bm s^{(1)}$, the corresponding erasure patterns for all row partitions after the swaps are correctable by $\mathcal{C}$ (see Section~\ref{sec:Sketch}).}
Note that we have performed  $\sum_{j=1}^{L} \mathcal |P_j|=n-k-\bar r$ swaps from the columns of $\tilde{\bm E}$ with coordinates in the set  $\cup_{j=1}^{L} \mathcal P_j$ to the $j'$-th column of $\bm Z$. Thus, each column in $\cup_{j=1}^{L} \mathcal P_j$ has column weight $n-k+\bar r-1$ and the $(n_{\mathsf c}L+j')$-th column has column weight $n-k-\bar r+\bar r=n-k$. Letting $j'=j'+1$ and repeating the above procedure $\bar r$ times ensures $\bm E$ to be $(n-k)$-regular.
%\item[b)] \textbf{Swapping elements between $\tilde{\bm E}$ and $\bm Z$.} We consider the $j'$-th column of $\bm Z$,
%  where $j'\in\Nat{\bar r}$. Without loss of generality, let $\mathcal E_{j'}=\{\vect{e}'_1,\ldots,\vect{e}'_\tau\}$,
%  where $\tau=n-k-\bar r\geq 0$, be a set of rows from the matrix $(\tilde{\bm E}\mid\bm Z)$. Note that from \cref{Def:
%    OptLRCs}, the total number of parities in \eqref{Eq: H_opt_LRC} is equal to $n-k$. Therefore, we have
%  \begin{align}
%    \label{total-parities} 
%    (\delta-1)L_\mathsf{c}+(L-L_\mathsf{c})n_\mathsf{c}+\bar{r}=n-k      
%  \end{align} 
%  and $\tau=n-k-\bar r\geq 0$.
%  
%  For $s\in\Nat{\tau}$, the $s$-th row in $\set{E}_{j'}$ will have a unique coordinate $p_s\in\chi(\bm e'_s)$ such that
%  $p_s\in\bigcup_{i=1}^{L}\mathcal P_i$. Note that $p_s\neq p_{s'}$, $s'\in\Nat{\tau}$, $s\neq s'$. Now, we begin
%  swapping elements in the set $\mathcal E_{j'}$. That is to say, swap values between $\bm e'_{s,p_s}$ and
%  $\bm e'_{s,n_\mathsf{c} L+j'}$. This process is repeated recursively for all $j'$. In each iteration, the weight of the
%  $p_s$-th column in $\bm E$ is reduced by $1$, and the weight of the $(n_\mathsf{c} L+j')$-th column is increased by
%  $\tau$, resulting in a column weight of $\tau+\bar r=n-k$.
\end{enumerate}
This completes the construction of $\bm E$, which has row and column weight $n-k$. In the following theorem, we show
that each row of $\bm E$ (considered as an erasure pattern) can be corrected by any code from the class of 
  distance-optimal $(r,\delta)$ information locality codes whose parity-check matrices are as in \cref{Eq: H_opt_LRC}
and are compliant with \cref{Eq: H_MDS}. Thus, this class of codes is MDS-PIR capacity-achieving.
\begin{theorem}
  \label{Th: LRCcap_proof}
  An $[n,k]$ distance-optimal $(r,\delta)$ information locality code $\code{C}$ with parity-check matrix as in
  \cref{Eq: H_opt_LRC} and satisfying \cref{Eq: H_MDS} is an MDS-PIR capacity-achieving code.
\end{theorem}
\begin{IEEEproof}
A sketch of the proof is given in Section~\ref{sec:Sketch}.
%  The proof is given in the extended version of the paper \cite[App.~F]{KumarLinRosnesGraell17_1sub}.
\end{IEEEproof}

In the following, we present an example to illustrate the construction of the matrix $\bm E$. 
%The existence of such a matrix ensures that the PIR protocols presented in \cref{sec:file-dep-PIR,sec:file-indep-PIR} achieve the finite PIR capacity $\const{C}_f$ \eqref{eq:PIRcapacity} and the asymptotic PIR capacity $\const{C}_\infty$ \eqref{eq:PIRasympt-capacity}, respectively.

\begin{example}
  Consider an $[n=7,k=4]$ Pyramid code $\mathcal C$ that is constructed from an $[n'=6,4]$ Reed-Solomon code over $\GF(2^3)$ with
  parity-check matrices
  \begin{align*}
      \bm H^{\mathcal C}&=\left(\begin{matrix}
          z^3 & 1 & 1 & 0   & 0   & 0 & 0\\
          0   & 0 & 0 & z^3 & z   & 1 & 0\\
          z^4 & 1 & 0 & z^5 & z^5 & 0 & 1
        \end{matrix}\right)
  \end{align*}
  and
   \begin{align*} 
                \bm H^{\mathsf{MDS}}&=\left(\begin{matrix}
          z^3 & 1 & z^3 & z & 1 & 0\\
          z^4 & 1 & z^5 & z^5 & 0 & 1
        \end{matrix}\right),
  \end{align*}
respectively, where $z$ denotes a primitive element of $\GF(2^3)$. It is easy to see that $\code{C}$ is a distance-optimal
  $(r=2,\delta=2)$ information locality code. We have $n_\mathsf{c}=3$, $L=L_{\mathsf c}=2$, and
  $\bar r\eqdef n\bmod n_\mathsf{c}=1$. Since $\rho_1=2$ and $\rho_2=1$, we get
  \begin{align*}
    \tilde{\bm E}=\left(\begin{matrix}
        \bm \pi_1 & \bm \pi_2\\
        \bm \pi_2 & \bm \pi_1
      \end{matrix}\right)=\left(\begin{array}{ccc|ccc}
                                  1 & 1 & 0 & 1 & 0 & 0\\
                                  0 & 1 & 1 & 0 & 1 & 0\\
                                  1 & 0 & 1 & 0 & 0 & 1\\
                                  \hline
                                  1 & 0 & 0 & 1 & 1 & 0\\
                                  0 & 1 & 0 & 0 & 1 & 1\\
                                  0 & 0 & 1 & 1 & 0 & 1\\
                                \end{array}\right), \;\bm Z=\left(\begin{matrix}
                                  0\\
                                  0\\
                                  0\\
                                  0\\
                                  0\\
                                  0
                                \end{matrix}\right),
  \end{align*}
  where $\bm \pi_1$ is a  $2$-regular $3 \times 3$ matrix and $\bm \pi_2$ is picked as the identity matrix. 
  The set of parity coordinates is $\mathcal P=\{3,6,7\}$, and we set $e_{1,3}^{(3)}=e_{1,6}^{(3)}=e_{1,7}^{(3)}=1$. As such,
  we get
  \begin{align*}
    \bm W=\left(\begin{matrix}
        0 & 0 & 1 & 0 & 0 & 1
      \end{matrix}\right)\text{ and }
                            \bm O=\left(\begin{matrix}
                                1
                              \end{matrix}\right).
  \end{align*}
  This completes Step a) of the construction above. Note that each row of $\bm E$ has now weight $3$. The second step of the procedure (Step b)) is as follows. Consider the first iteration, $j'=1$. In the first row partition we choose $\bm s^{(1)}=(s^{(1)}_1=1,s^{(1)}_2=0)$. %Furthermore, $s^{(1)}_1+s^{(1)}_2=1$. 
  Taking $\mathcal R_1^{(1)}=\{2\}$, we do the swap between the coordinates $(i'=2,p=3\in\mathcal P_1)$ and $(i',6+j')$. For the second row partition we have $\bm s^{(2)}=(0,1)$ which is a right cyclic shift of $\bm s^{(1)}$. Taking $\mathcal R_2^{(2)}=\{6\}$, we do the swap between the coordinates $(i'=6,p=6\in\mathcal P_2)$ and $(i',6+j')$. Thus, we have 
\begin{align*}
\begin{split}
	e_{2,3}^{(1)}=0, \;e_{2,7}^{(1)}=1,\\
	e_{3,6}^{(2)}=0, \;e_{3,7}^{(2)}=1.
\end{split}
\end{align*}
Since $\bar r=1$, this completes Step b), which results in
\begin{align*}
  \bm E=\left(\begin{array}{ccc|ccc|c}
                1 & 1 &   0 & 1 & 0 & 0 &   0\\
                0 & 1 &\r 0 & 0 & 1 & 0 &\r 1\\
                1 & 0 &   1 & 0 & 0 & 1 &   0\\
                \hline
                1 & 0 & 0 & 1 & 1 &   0&   0\\
                0 & 1 & 0 & 0 & 1 &   1&   0\\
                0 & 0 & 1 & 1 & 0 &\r 0&\r 1\\
                \hline
                0 & 0 & 1 & 0 & 0 & 1 & 1
              \end{array}\right).
\end{align*}
The entries in red indicate the swapped values within each row. It can easily be verified that each row of $\bm E$ is an
erasure pattern that is correctable by code $\mathcal C$.
\end{example}

\section{Sketch of Proof of \cref{Th: LRCcap_proof}}
\label{sec:Sketch}

In the following, we give a sketch of the proof of \cref{Th: LRCcap_proof}. A more detailed proof is presented in \cite[App.~F]{KumarLinRosnesGraell17_1sub}. 
According to Theorem~\ref{Th:CapAch}, to prove that a distance-optimal $(r,\delta)$ information locality code $\mathcal C$ is MDS-PIR capacity-achieving, it is sufficient to prove that there exists an $(n-k)$-regular matrix $\bm E$ whose rows represent erasure patterns that are correctable by $\mathcal C$. The construction of such a matrix $\bm E$, provided in Section~\ref{Sec: Ematrix_LRC}, involves two steps as follows.
\begin{itemize}
\item[a)] The submatrices $\tilde{\bm E}$, $\bm W$, $\bm Z$, and $\bm O$ are systematically constructed such that the row weight constraint is satisfied. 
\item[b)] Swap elements in certain rows of matrices $\tilde{\bm E}$ and $\bm Z$ in order to meet the column weight constraint of $\bm E$. 
\end{itemize}

The proof is a two-step procedure. First, we prove that all rows in $\bm E$ after Step a) are correctable by $\mathcal C$. Secondly, we prove that the swaps in certain rows in Step b) ensure that the resulting rows are correctable erasure patterns. We will make use of the following lemma.
\begin{lemma}
  \label{Lem: Capacity_LRC}
  Let $\mathcal C$ be an $[n,k]$ distance-optimal $(r,\delta)$ information locality code consisting of ${L_{\mathsf c}}$ local codes and with  parity-check   matrix as in \cref{Eq: H_opt_LRC}. Additionally, it adheres to the condition in \cref{Eq: H_MDS}. Then, $\mathcal C$ can   simultaneously correct $\delta-1+\nu_j$ erasures, $\nu_j\geq0$, in each local code $\code{C}|_{\set{S}_j}$ provided that the number of   global parities available is at least $\nu_1+\cdots+\nu_{L_{\mathsf{c}}}$.
\end{lemma}
\begin{IEEEproof}
The proof is given in  \cite[App.~F]{KumarLinRosnesGraell17_1sub}.	
\end{IEEEproof}

Consider the erasure patterns in the first row partition of $\bm E$ after Step a). Each of these patterns has $\nu_j=\rho_j-(\delta-1)$, $j\in\Nat{L_{\mathsf{c}}}$, erasures occurring in the coordinates corresponding to the local code $\code{C}|_{\set{S}_j}$ that cannot be corrected locally. Furthermore, the number of nonerased global parities is equal to $\gamma_{\mathsf{tot}}+\bar{r}$, where $\gamma_{\mathsf{tot}}$ is the total number of nonerased global parity coordinates present in the column partitions $L_{\mathsf c}+1,\ldots,L$. It can be shown that $\sum_{j=1}^{L_{\mathsf c}}\nu_j\leq \gamma_{\mathsf{tot}}+\bar{r}$ (see \cite[proof of Lem.~8]{KumarLinRosnesGraell17_1sub}). From \cref{Lem: Capacity_LRC}, all erasures in the $L_{\mathsf c}$ local codes are correctable. This enables the code to correct the remaining erasures at the coordinates of $\mathcal C$ in the set $\cup_{j=L_{\mathsf c}+1}^{L} \mathcal P_j$. Thus, the erasure patterns in the first row partition of $\bm E$ after Step a) are correctable. Through induction, one can prove that the erasure patterns in the remaining $L-1$ row partitions are also correctable. The erasure patterns in $(\bm W|\bm O)$ are correctable by $\mathcal C$ as they pertain to the local and global parity symbols. This completes the first part of the proof.

We now address the second part of the proof. Note that the columns with coordinates in $\mathcal P_j$, $j\in\Nat{L}$, have column weight $n-k+\bar{r}$ after Step a). Step b) involves the swapping of one entries from these coordinates with zero entries in the column coordinates of $\bm Z$. 
% In particular, due to the regularity of $\tilde{\bm E}$, we need to swap $\bar r(\delta-1)$ `$1$' in the $j$-th row
% partition across the $L$ column partitions in the $j$-th column partition across the $L$ row partitions of
% $\tilde{\bm E}$.  to `$0$'
The swapping is done to ensure that the column weight of the columns indexed by $\mathcal P_j$, $j\in\Nat{L}$, is reduced to $n-k$, while those of the columns of $\bm Z$ are increased to $n-k-\bar{r}$. Since $\bm O$ is an all-one matrix, the columns of $\bm E$ with indices in $\mathcal P_{L+1}$ have also weight $n-k$. It is possible to show that such a swapping always exists. Overall, the resulting matrix $\bm E$ is $(n-k)$-column regular. To ensure that the erasure patterns are correctable, we use \cref{Lem: Capacity_LRC}. For each row, 
\begin{align}
\label{Eq: CountingArg}
	\sum_{j=1}^{L_{\mathsf c}}\nu_j\leq \gamma_{\mathsf{tot}}+   \modify{\gamma_{L+1}},%\bar{r}.
\end{align} 
\modify{where $\gamma_{L+1}$ is number of nonerased parity coordinates in column partition $L+1$, must hold.} 
Clearly, if for a certain row of $(\tilde{\bm E}\mid\bm Z)$ a one from a column from a column partition in $\Nat{L_{\mathsf c}+1:L}$ (corresponding to $\tilde{\bm E}$) is swapped with a zero in a column from partition $L+1$ (corresponding to $\bm Z$), then the resulting erasure pattern is still correctable by $\mathcal C$ as \cref{Eq: CountingArg} is still valid. On the other hand, for $j\in\Nat{L_{\mathsf c}}$, if for a certain row of $(\tilde{\bm E}\mid\bm Z)$ a one from the $j$-th column partition is swapped with a zero  in the $(L+1)$-th column partition, then such a row is still a correctable erasure pattern provided that $\nu_j > 0$ before the swap. This is easy to see as the swapping procedure reduces $\nu_j$ and  $\modify{\gamma_{L+1}}$ by one. Thus, \cref{Eq: CountingArg} is still satisfied. From the aforementioned arguments and the fact that each row of any row partition of $(\tilde{\bm E}\mid\bm Z)$ has at most $\bar r$ swaps of ones occurring from the set of $\Nat{L}$ column partitions and zeroes from the $(L+1)$-th partition,  \modify{it follows that the swaps according to Step b)  are valid over all $\bar{r}$ iterations  (valid in the sense that the resulting erasure patterns are correctable by $\mathcal{C}$) if }
\modify{
\begin{align}
\label{Eq: SwapCond}
	\sum_{j=1}^{L_{\mathsf c}}\nu_j+\sum_{j=L_{\mathsf c}+1}^{L}(m-(\delta-1))\geq\bar r.
\end{align}
This is a counting argument, where according to Step b) {\textcolor{black}{for each row}} we restrict swapping $\nu_j$ coordinates in the $j$-th column partition, $j\in\Nat{L_{\mathsf c}}$, and $m-(\delta-1)$ coordinates in the column partitions $\Nat{L_{\mathsf c}+1:L}$ to make sure (following the arguments above) that the resulting erasure pattern after the swap is correctable by $\mathcal{C}$.  Using that $\nu_j=\rho_j-(\delta-1)$ and $t=n-k-mL$, it can be shown that the left hand side of  \eqref{Eq: SwapCond} can be lowerbounded by %where $\rho_j\in\Nat{m:m_1}$, the left hand side of  \eqref{Eq: SwapCond} can be lowerbounded by  % when $\nu_j=m-(\delta-1)$ that occurs when $n-k\mid L$. Thus, we have the lower bound as
%\begin{align}
%\label{Eq: LHS_Reduce}
%\begin{split}
%\sum_{j=1}^{L}(m-(\delta-1)) %L(m-(\delta-1))
 $n-k-L(\delta-1)$ 
 when $t \leq L_{\mathsf c}$. Setting  $n=\bar r+L(r+\delta-1)$ and $k=L_{\mathsf c}r$, it follows that \eqref{Eq: SwapCond}   reduces to $L\geq L_{\mathsf c}$. 
 %From \cref{Eq: LHS_Reduce}, $k=L_{\mathsf c}r$, and $\bar r=n-L(r+\delta-1)$, \cref{Eq: SwapCond} reduces to $L\geq L_{\mathsf c}$.
By definition, this is always true. When $t > L_{\mathsf c}$, the left hand side of  \eqref{Eq: SwapCond}  is equal to $n-k-L(\delta-1) + L_{\mathsf c} - t$, and it can be shown that this is always larger than or equal to $\bar{r}$, since $t \leq L$ (details omitted for brevity). \modify{It follows that for all $\bar{r}$ iterations  and for all row partitions in the systematic procedure in Step b) there exists a valid swap such that  the resulting erasure patterns are still correctable by $\mathcal C$.}}

\section{Conclusion}
%In \cite{KumarRosnesGraell17_1}, the authors showed numerically that their PIR protocol was able to achieve the highest PIR rate when the underlying code was a Pyramid code. In fact, this rate was equal to the MDS-PIR capacity when the number of files tends towards infinity. In our extended version of this paper, we show the conditions required by our protocols to achieve the MDS-PIR capacity given any underlying linear code. We refer to such codes as MDS-PIR capacity-achieving codes. Interestingly, Pyramid codes are a part of a larger class of distance optimal LRCs that perform efficient node repair in DSSs. 

We formally proved that a class of distance-optimal LRCs, an important class of codes used in DSSs, are MDS-PIR capacity-achieving codes. %This particular class of codes includes Pyramid codes and other constructions given in the literature.
 The considered class of codes  includes Pyramid codes and other constructions of LRCs given in the literature.
 
 %The proof involves constructing PIR capacity-achieving matrices for such LRCs. %Knowledge of this matrix allows to construct the MDS-PIR capacity-achieving protocols that is presented in our extended version.

%
%
%We generalized the PIR protocol proposed in \cite{taj16} for a DSS with a single spy node and where data is stored using an MDS code to the case where an arbitrary systematic linear code of rate $R>1/2$ is used to store data. We also presented an algorithm to optimize the cPoP of the protocol. The optimization leads to a cPoP close to its theoretical lower bound. Interestingly, for certain codes, the lower bound on the cPoP can be achieved.

\balance

\bibliographystyle{IEEEtran}
{\small \itemsep 10ex
\bibliography{Bib_PIR.bib,defshort1.bib,biblio1.bib}
}

\end{document}